# On Determining the Spectrum of Primordial Inhomogeneity from the $COBE$[1] DMR Sky Maps: I. Method

Krzysztof M. Górski[2,3,4]

## ABSTRACT

The natural approach to a spectral analysis of data distributed on the sky employs spherical harmonic decomposition. A common problem encountered in practical astronomy is the lack of full sky coverage in the available data. For example, the removal of the galactic plane data from the $COBE$ DMR sky maps compromises Fourier analysis of the cosmic microwave background (CMB) temperature distribution due to the loss of orthogonality of the spherical harmonics. An explicit method for constructing orthonormal functions on an incomplete (e.g. Galaxy-cut) sphere is presented. These functions should be used in the proper Fourier analysis of the $COBE$ DMR sky maps to provide the correct input for the determination of the spectrum of primordial inhomogeneity. The results of such an analysis are presented in an accompanying *Letter*. A similar algebraic construction of appropriate functions can be devised for other astronomical applications.

*Subject headings*: Cosmic microwave background, Methods: Analytical

# 1. INTRODUCTION

Analysis of the first year of *COBE* DMR mission data resulted in the discovery of temperature anisotropy on the microwave sky of cosmological origin (Smoot *et al.* 1992, Bennett *et al.* 1992, Wright *et al.* 1992). A summary of the results from two years of observations was presented in Bennett *et al.* (1994). The *COBE* DMR instrument has observed the microwave sky at three frequencies (31.5, 53, and 90 GHz) with pairs of 7° (FWHM) antennas separated by 60°. The temperature difference measurements were used to construct whole sky anisotropy maps (Janssen & Gulkis 1992) binned into 6144 equal area ($\sim 2°\!.6 \times 2°\!.6$) pixels on a quadrilateralized spherical cube projection (White & Stemwedel 1992). The temperature fluctuation assigned to a given pixel $p$, $\Delta(p) \equiv \delta T/T(\theta_p, \phi_p)$, is primarily due to cosmic microwave background (CMB) radiation, emission from the Galaxy, and receiver noise. Systematic errors in the measurements have been carefully studied (Kogut *et al.* 1992, Bennett *et al.* 1994), and are ignored in the present analysis.

The observed full sky temperature distribution, $\Delta(p) = \Delta_{CMB}(p) + \Delta_{Galaxy}(p) + \Delta_{Noise}(p)$, can be expanded in spherical harmonics

$$\Delta(p) = \sum_{\ell=0}^{\ell_{max}} \sum_{m=-\ell}^{\ell} \left\{ \left[ a_{\ell m}^{CMB} + a_{\ell m}^{Galaxy} \right] w_{\ell}^{DMR} + a_{\ell m}^{Noise} \right\} w_{\ell}^{pix} Y_{\ell m}(\theta_p, \phi_p). \qquad (1)$$

Here, the $Y_{\ell m}$ are real, orthonormal spherical harmonics†, $w_{\ell}^{DMR}$ are the filter coefficients of the DMR beam pattern (Wright *et al.* 1994), and $w_{\ell}^{pix}$ are the filter coefficients used to model the extra smoothing due to pixelization of the data (actually a circular top-hat window of solid angle $\Omega_{pix} = 4\pi/6144$). $a_{\ell m}$ are harmonic coefficients of the CMB anisotropy, galactic emission, and instrument noise. In the sum over $\ell$, the lower limit is taken deliberately at $\ell = 0$; although the instrument does not directly measure the absolute mean temperature on the sky, both $a_{00}^{Galaxy}$ and an arbitrary (but small) offset introduced by the map making algorithm exist. Similarly, the *a priori* CMB dipole anisotropy removal employed in the two year data analysis (Bennett *et al.* 1994) is only approximate and leaves a small residual $\ell = 1$ component in the maps. Due to the pixel size, a formal upper limit in the sum over $\ell$ is near $\ell_{max} \sim 100$ for the raw, noise-dominated maps. However, the 7° FWHM DMR antennas are sensitive to sky signals only up to $\ell \sim 30$, so that at $\ell \gtrsim 30$ $\Delta(p)$ in eq.(1) is, in practice, dominated entirely by the $a_{\ell m}^{Noise}$ term.

---

† $Y_{\ell m}(\theta, \phi) = \sqrt{\frac{2\ell+1}{2}} \sqrt{\frac{(\ell-|m|)!}{(\ell+|m|)!}} P_{\ell}^{|m|}(\cos\theta) f(\phi)$, where $f(\phi) = ((2\pi)^{-1/2}, \pi^{-1/2}\cos m\phi$, or $\pi^{-1/2}\sin|m|\phi)$, for $m = 0, > 0$, or $< 0$. $P_{\ell}^{m}(x) \overset{m \geq 0}{=} (-1)^m (1-x^2)^{m/2} \frac{d^m}{dx^m} P_{\ell}(x)$, where $P_{\ell}(x) = \frac{1}{2^{\ell}\ell!} \frac{d^{\ell}}{dx^{\ell}} (x^2 - 1)^{\ell}$ are Legendre polynomials.

A major aim of the DMR data analysis is to extract the cosmological signal $\Delta_{CMB}(p)$, or the coefficients $a_{\ell m}^{CMB}$ from the sky maps. To achieve this, one needs to identify and separate the signals due to receiver noise and galactic emission. The receiver noise in a given map can be well characterized statistically using the known properties of the radiometer and the sky coverage. The dominant galactic emission can be eliminated most easily by removing all pixels within 20 degrees of the galactic plane. A more detailed approach is to model the galactic emission in detail utilizing the multifrequency nature of the data (Bennett *et al.* 1992). The results of such modelling indicate that galactic emission at high latitudes is relatively weak, while the emission in the plane cannot be removed accurately enough to warrant using the data for anisotropy studies. Thus the technique of cutting the plane and ignoring the high-latitude emission appears to yield the best combination of minimizing galactic emission while maximizing the signal to noise ratio, since the modelling technique includes the relatively noisy 31.5 GHz DMR data.

Implementing a galactic cut in the data creates a problem for Fourier analysis of the sky maps, since the drastically reduced sky coverage (e.g., a $|b| = 20°$ Galaxy cut results in $\sim 35\%$ loss in the solid angle) does not provide support for the orthogonality of the spherical harmonics (eq. 1).

In this *Letter*, a method for generating an orthonormal basis of functions on the pixelized, Galaxy-cut *COBE* DMR sky is presented. This method is a practical implementation of the Gram-Schmidt orthogonalization (see e.g. Kostrikin & Manin 1989) of a set of linearly independent vectors. These algebraically constructed orthonormal basis functions are used for (1) a proper Fourier analysis of the reduced coverage sky maps, (2) a derivation of the correlation matrices of Fourier amplitudes for the model anisotropy and the instrumental noise pattern, and (3) a computation of the likelihood function for the cosmological structure formation model parameters to be inferred from the experimental data.

Hereafter: bold, upper case letters denote matrices; bold, lower case letters denote vectors; subscripts $i$, $j$, $k$, $l$ denote indices of orthonormal functions; and indices or subscripts $p$, $q$ are pixel labels, which identify the angular position on the sky.

## 2. ORTHONORMAL FUNCTIONS AND FOURIER ANALYSIS ON AN INCOMPLETE SPHERE

Consider a linear space $\mathcal{F}_{\ell_{max}}$ of bounded functions on a sphere spanned by spherical harmonics up to order $\ell_{max}$. Let us form an $(\ell_{max}+1)^2$-dimensional vector $\mathbf{y} = (Y_{0,0}, Y_{1,-1}, Y_{1,0}, Y_{1,1}, Y_{2,-2}, Y_{2,-1}, \ldots, Y_{\ell_{max},\ell_{max}})$. An index pair, $(\ell, m)$, of the spherical harmonic degree and order is mapped into $i = \ell^2 + \ell + 1 + m$, a single index of the vector $y_i$. The inverse mapping is given by $\ell = \text{integer}\,(\sqrt{i-1})$, and $m = i - (\ell^2 + \ell + 1)$. A function $f \in \mathcal{F}_{\ell_{max}}$ is defined as a finite Fourier series $f(\Omega) = \sum_{i=1}^{(\ell_{max}+1)^2} a_i\, y_i(\Omega) \equiv \mathbf{a}^T \cdot \mathbf{y}$, where $a_i = \int_{4\pi} d\Omega\, f(\Omega)\, y_i(\Omega)$. For any functions $f$, $g \in \mathcal{F}_{\ell_{max}}$ define the scalar product specific to the pixelized sky:

$$\langle f\, g \rangle_{\{full\ sky\}} \equiv \Omega_{pix} \sum_{p \in \{full\ sky\}} f(p)\, g(p) \approx \int_{4\pi} d\Omega\, f(\Omega)\, g(\Omega), \tag{2}$$

where $\{full\ sky\}$ denotes the entire set of 6144 sky map pixels. Spherical harmonics are orthonormal in the actual DMR pixelized sky scalar product, $\langle \mathbf{y} \cdot \mathbf{y}^T \rangle_{\{full\ sky\}} = \mathbf{I}$, to an absolute accuracy better than $10^{-3}$ for $\ell \leq 30$.

When a Galaxy cut is applied to the sky map the set of pixels that enter the scalar product is reduced. An explicit notation for this is

$$\langle f\, g \rangle_{\{cut\ sky\}} = \Omega_{pix} \sum_{p \in \{cut\ sky\}} f(p)\, g(p). \tag{3}$$

For example, the Galaxy cut at $|b| = 20°$ reduces the number of pixels from 6144 to 4016. If only a subset of a sphere provides support for the scalar product, then the spherical harmonics are linearly independent due to their different functional forms, but are not orthogonal. A new functional basis, orthonormal in the scalar product of eq. (3), can be algebraically constructed from the functions y as follows.

Evaluate the coupling matrix for the spherical harmonics on the cut sky

$$\left\langle \tilde{\mathbf{y}} \cdot \tilde{\mathbf{y}}^T \right\rangle_{\{cut\ sky\}} \equiv \mathbf{W}. \tag{4}$$

Here, $\tilde{\mathbf{y}}$ are spherical harmonics with amplitudes slightly suppressed by the factor $w^{pix}_{\ell(i)}$ (e.g. $w^{pix}_{30} \sim 0.93$), due to pixelization. The Kowalewski-Gram determinant det $\mathbf{W}$ (Sansone 1959), is an indicator of whether a set of underlying functions is linearly independent. In the case of reduced sky coverage, $\mathbf{W}$ is a non-diagonal, symmetric, non-singular, positive definite matrix. Hence, $\mathbf{W}$ can be Choleski-decomposed into the product of a lower triangular matrix $\mathbf{L}$ and its transpose

$$\mathbf{W} = \mathbf{L} \cdot \mathbf{L}^T. \tag{5}$$

The essential point of this method relates to the observation that Choleski decomposition of the coupling matrix $\mathbf{W}$ provides a direct means of Gram-Schmidt orthogonalization.

Compute the inverse matrix $\mathbf{\Gamma} = \mathbf{L}^{-1}$, and define a new set of functions on the cut sky

$$\boldsymbol{\psi}(p) = \mathbf{\Gamma} \cdot \tilde{\mathbf{y}}(p). \tag{6}$$

By construction

$$\left\langle \boldsymbol{\psi} \cdot \boldsymbol{\psi}^T \right\rangle_{\{cut\ sky\}} = \mathbf{\Gamma} \cdot \left\langle \tilde{\mathbf{y}} \cdot \tilde{\mathbf{y}}^T \right\rangle_{\{cut\ sky\}} \cdot \mathbf{\Gamma}^T = \mathbf{\Gamma} \cdot \mathbf{L} \cdot \mathbf{L}^T \cdot \mathbf{\Gamma}^T = \mathbf{I}. \tag{7}$$

Hence, the functions $\boldsymbol{\psi}$ are orthonormal on the pixelized, cut sky, and are suitable for the Fourier analysis of functions in $\mathcal{F}_{\ell_{max}}$ with support restricted to the cut sky. This algorithm has been numerically implemented for $\ell_{max} = 30$. Orthonormality of the resulting 961 $\boldsymbol{\psi}$ functions was achieved to an absolute accuracy better than $10^{-7}$.

Using the new basis $\boldsymbol{\psi}$, a function $f(p) = \mathbf{a}^T \cdot \mathbf{y}(p)$ can be Fourier decomposed in the cut sky:

$$f(p)|_{p \in \{cut\ sky\}} = \mathbf{c}^T \cdot \boldsymbol{\psi}(p), \quad \text{where} \quad \mathbf{c} = \langle f\,\boldsymbol{\psi}\,\rangle_{\{cut\ sky\}}. \qquad (8)$$

It follows from eq.(6) that the relations between expansion coefficients in the $\mathbf{y}$ and $\boldsymbol{\psi}$ bases are

$$\mathbf{c} = \mathbf{L}^T \cdot \mathbf{a}, \quad \text{and} \quad \mathbf{a} = \boldsymbol{\Gamma}^T \cdot \mathbf{c}. \qquad (9)$$

It is important to note that due to the triangular form of $\mathbf{L}$ and $\boldsymbol{\Gamma}$, each coefficient $c_i$ is a combination of coefficients $a_j$ of order $j \geq i$. Hence, truncating the first four components of the vector $\mathbf{c}$ unambiguously removes the monopole and dipole degrees of freedom from the temperature map, but does not affect the information content of the map for higher orders of $\ell$. This property is a consequence of the specific ordering of the $\mathbf{y}$ basis.

## 3. MODEL CMB ANISOTROPY CORRELATIONS IN THE $\boldsymbol{\psi}$ BASIS

Mainstream cosmological structure formation models invoke homogeneous, isotropic, Gaussian fields of small amplitude curvature perturbations superposed on a flat background, specified by the spatial correlation function, or its Fourier transform, the power spectrum. The angular distribution of CMB temperature anisotropy induced by such curvature perturbations (Sachs & Wolfe 1967) is easily described in Fourier language (e.g. Peebles 1981). Individual spherical harmonic coefficients of such a CMB anisotropy field are Gaussian-distributed in the theoretical ensemble of initial conditions for structure formation. The variances of the probability distributions of individual modes, $\langle a^2_{i(\ell,m)} \rangle$, are uniquely expressed as integrals over the power spectrum, and depend only on $\ell$ due to the statistical isotropy of the CMB temperature field. Moreover, the ensemble correlation matrix of the coefficients $\mathbf{a}_{CMB}$ is diagonal: $\langle \mathbf{a}_{CMB} \cdot \mathbf{a}^T_{CMB} \rangle = \text{diag}\left\{ \langle a^2_{i(\ell,m)} \rangle \right\}$. Using eq.(9) one readily obtains the correlation matrix for the Fourier coefficients of the theoretical cut sky CMB anisotropy

$$\mathbf{C}_{CMB} \equiv \langle \mathbf{c}_{CMB} \cdot \mathbf{c}^T_{CMB} \rangle = \mathbf{L}^T \cdot \langle \tilde{\mathbf{a}}_{CMB} \cdot \tilde{\mathbf{a}}^T_{CMB} \rangle \cdot \mathbf{L}, \qquad (10)$$

where the effect of DMR beam smearing is included in the vector $\tilde{\mathbf{a}}_{CMB}$ with elements $\tilde{a}_i^{CMB} = a_i^{CMB} w_{\ell(i)}^{DMR}$, and pixelization smoothing is included in the matrix $\mathbf{L}$ (eqs. 4, 5). The Gaussian ensemble probability density of $\mathbf{c}_{CMB}$ is specified by the matrix $\mathbf{C}_{CMB}$:

$$P(\mathbf{c}_{CMB})\, d\mathbf{c}_{CMB} \propto d\mathbf{c}_{CMB}\, \exp\left(-\mathbf{c}_{CMB}^T \cdot \mathbf{C}_{CMB}^{-1} \cdot \mathbf{c}_{CMB}/2\right) / \sqrt{\det(\mathbf{C}_{CMB})}.$$

In the long wavelength perturbation regime probed by $COBE$ DMR, the power spectrum of density perturbations is usually assumed to be a power law, $P(k) \propto k^n$. From this and previous assumptions, one can derive a compact formula for the expectation values of theoretically predicted CMB anisotropy multipole amplitudes (see Bond & Efstathiou 1987, and Fabbri, Lucchin, & Matarrese 1987)

$$\left\langle a_{i(\ell,m)}^2 \right\rangle = a_2^2\, \frac{\Gamma\left(\ell + \frac{n-1}{2}\right)\, \Gamma\left(\frac{9-n}{2}\right)}{\Gamma\left(\ell + \frac{5-n}{2}\right)\, \Gamma\left(\frac{3+n}{2}\right)}, \tag{11}$$

where $a_2$ is the rms quadrupole coefficient, often specified as $Q_{rms-PS} = \sqrt{\frac{5}{4\pi}} a_2$, and $\Gamma$ is the gamma function, not to be confused with the matrix $\mathbf{\Gamma}$. A broad category of inflationary structure formation models predicts $n \sim 1$, the Harrison-Zel'dovich spectrum. Using existing observational data to test the hypothesis that eq.(11) provides a satisfactory description of the universe on large scales is a critically important aspect of contemporary cosmology.

## 4. NOISE CORRELATIONS IN THE $\psi$ BASIS

The effect of radiometer noise on the sky maps constructed from the $COBE$ DMR observations can be described to leading order (i.e. neglecting the small noise correlations between pixels separated by 60°, see Lineweaver et al. 1994) as a spatially uncorrelated Gaussian process of vanishing mean, $\langle \Delta_N(p) \rangle_{\{noise\ ensemble\}} = 0$, and pixel dependent variance $\langle \Delta_N(p) \Delta_N(q) \rangle_{\{noise\ ensemble\}} = \sigma^2(p)\, \delta_{pq}$. The function $\sigma^2(p)$ is determined from the known noise characteristics of each individual radiometer channel and the number of observations per pixel. $\sigma(p)$ varies with both frequency and channel, and combinations thereof, and can be determined to a few percent accuracy.

Fourier decomposition of the random noise pattern on the cut sky yields

$$\Delta_N(p)|_{p \in \{cut\ sky\}} = \mathbf{c}_N^T \cdot \boldsymbol{\psi}(p), \quad \text{where} \quad \mathbf{c}_N = \langle \Delta_N \boldsymbol{\psi} \rangle_{\{cut\ sky\}}. \quad (12)$$

On averaging over an ensemble of Gaussian noise realizations, $\langle \mathbf{c}_N \rangle_{\{noise\ ensemble\}} = 0$, and the correlation matrix of the Fourier components of the noise pattern is

$$\begin{aligned}
\mathbf{C}_N &= \langle \mathbf{c}_N \cdot \mathbf{c}_N^T \rangle_{\{noise\ ensemble\}} \\
&= \langle \langle \Delta_N(p) \Delta_N(q) \rangle_{\{noise\ ensemble\}} \boldsymbol{\psi}(p) \cdot \boldsymbol{\psi}^T(q) \rangle_{p,q \in \{cut\ sky\}} \\
&= \Omega_{pix} \langle \sigma^2(p) \boldsymbol{\psi}(p) \cdot \boldsymbol{\psi}^T(p) \rangle_{p \in \{cut\ sky\}}.
\end{aligned} \quad (13)$$

Analogously, it can be shown that

$$\langle c_i c_j c_k c_l \rangle_{\{noise\ ensemble\}} = \langle c_i c_j \rangle \langle c_k c_l \rangle + \langle c_i c_k \rangle \langle c_j c_l \rangle + \langle c_i c_l \rangle \langle c_j c_k \rangle, \quad (14)$$

and similarly for higher even moments, and that all odd moments are identically zero. Hence the probability distribution of the noise multipoles $\mathbf{c}_N$ is purely Gaussian, fully specified by the matrix $\mathbf{C}_N$: $P(\mathbf{c}_N) d\mathbf{c}_N \propto d\mathbf{c}_N \exp\left(-\mathbf{c}_N^T \cdot \mathbf{C}_N^{-1} \cdot \mathbf{c}_N / 2\right) / \sqrt{\det(\mathbf{C}_N)}$.

Spatial uniformity in the noise distribution, $\sigma(p) = const$, would result in a diagonal correlation matrix $\mathbf{C}_N$. The actual non-uniform noise distribution in the *COBE* DMR sky maps is specified by the strongly diagonally dominated, symmetric matrices $\mathbf{C}_N$.

## 5. LIKELIHOOD ANALYSIS

Since Gaussian statistical ensembles were invoked in §3 and §4 for the modelling of both cosmological and receiver noise temperature fluctuations in the cut sky maps, the *exact* probability distribution for the measurable Fourier amplitudes, $\hat{\mathbf{c}} = \hat{\mathbf{c}}_{CMB} + \hat{\mathbf{c}}_N$, of the noise contaminated cosmological CMB anisotropy on the cut sky is

$$P(\hat{\mathbf{c}}) d\hat{\mathbf{c}} = \frac{d\hat{\mathbf{c}}}{(2\pi)^{N/2}} \frac{e^{-\frac{1}{2} \hat{\mathbf{c}}^T \cdot (\mathbf{C}_{CMB} + \mathbf{C}_N)^{-1} \cdot \hat{\mathbf{c}}}}{\sqrt{\det(\mathbf{C}_{CMB} + \mathbf{C}_N)}}. \quad (15)$$

The monopole and dipole components of the sky maps, which are not physically relevant to the power spectrum estimation, can be exactly removed from the analysis by integrating over the first four components of the vector $\hat{\mathbf{c}}$. This operation removes the first four rows and columns of the correlation matrix, and reduces its rank to $N = (\ell_{max} + 1)^2 - 4$. The remaining $N$ elements of the vector $\hat{\mathbf{c}}$ retain the entire information content of the sky map within the spectral range $\ell \in [2, \ell_{max}]$. The need to choose a specific value for $\ell_{max}$ in a practical implementation of a Fourier decomposition of the sky maps creates the problem of leakage of signal from the spectral range $\ell > \ell_{max}$ into the $\ell \in [2, \ell_{max}]$ regime. Clearly, it is impossible in the analysis of an unfiltered sky map at a single frequency to separate correctly the true sky signal from the high-$\ell$ noise contamination. However, the definition of the probability distribution of the measurable Fourier amplitudes $\hat{\mathbf{c}}$ in eq. (15) can be naturally extended to a multifrequency analysis. For example, a simultaneous analysis of the DMR 53 GHz and 90 GHz data involves a composite correlation matrix of the form

$$\mathbf{C}_{53 \oplus 90} = \begin{pmatrix} \mathbf{C}_{CMB} + \mathbf{C}_{N53} & \mathbf{C}_{CMB} \\ \mathbf{C}_{CMB} & \mathbf{C}_{CMB} + \mathbf{C}_{N90} \end{pmatrix}, \qquad (16)$$

and a measured amplitude vector

$$\hat{\mathbf{c}}_{53 \oplus 90}^T = \left( \hat{\mathbf{c}}_{53}^T, \hat{\mathbf{c}}_{90}^T \right). \qquad (17)$$

The cross-correlation terms in eq.(16) limit the influence of the high-$\ell$ frequency dependent noise leakage (and, in fact, of any frequency dependent signal contamination due to the galactic foreground emission or systematic effects) on the inferred frequency independent cosmological signal. In a simultaneous analysis of multifrequency data sets full advantage is taken of both the auto-correlation structure of individual maps and the cross-correlation between different frequency maps. Thus, all of the available information from the multiple experimental data sets is utilized.

By assuming values for the cosmological parameters $a_2$ and $n$ (eq. 11), one can use eq.(15) in a frequentist Monte Carlo simulation of synthetic vectors $\mathbf{c}$. However, since we know the measured vector $\hat{\mathbf{c}}$ and want to know the likely values of the ensemble averaged power spectrum parameters $a_2$ and $n$, which specify the matrix $\mathbf{C}_{CMB}$, a Bayesian approach (e.g. Berger 1980) is suggested, wherein eq.(15) is a definition of the likelihood for the parameters $a_2$ and $n$, given the specific data set. This approach and subsequent analysis is discussed in an accompanying *Letter* (Górski *et al.* 1994) on the application of the formalism presented in this work.

## 6. SUMMARY

An algebraic method for constructing orthonormal basis functions on an incomplete sphere, designed for use in the Fourier analysis of the Galaxy-cut *COBE* DMR sky maps, has been presented. The method involves a completely general implementation of the Gram-Schmidt orthogonalization of linearly independent vectors, and could be easily modified for use in other astronomical projects with incomplete sky coverage.

The primordial power spectrum determination based on the likelihood method presented in §5 applied to the two year *COBE* DMR sky maps is presented in an accompanying *Letter* (Górski *et al.* 1994).

## REFERENCES

Bennett, C.L., *et al.* 1992, ApJ, 396, L7

Bennett, C.L., *et al.* 1994, *COBE* preprint No. 94-01, ApJ submitted

Berger, J.O. 1980, Statistical Decision Theory and Bayesian Analysis, (New York:Springer-Verlag)

Bond, J. R., & Efstathiou, G. 1987, MNRAS, 226, 655

Fabbri, R., Lucchin, F., & Matarrese, S. 1987, ApJ, 315, 1

Górski, K.M., *et al.* 1994, *COBE* preprint, ApJLetters submitted

Janssen, M.A., & Gulkis, S. 1992, in The Infrared and Submillimeter Sky after *COBE*, eds. M. Signore & C. Dupraz (Dordrecht:Kluwer)

Kogut, A., *et al.* 1992, ApJ, 401, 1

Kostrikin, A.I., & Manin, Yu.I. 1989, Linear Algebra and Geometry, (New York:Gordon and Breach Science Publishers)

Lineweaver, C. *et al.* 1994, $COBE$ preprint, ApJLetters submitted

Peebles, P.J.E. 1981, ApJ, 243, L119

Sachs, R. K., & Wolfe, A. M. 1967, ApJ, 147, 73

Sansone, G. 1959, Orthogonal Functions, (New York:Interscience Publishers)

Smoot, G. F. *et al.* 1992, ApJ, 396, L1

White, R.A., & Stemwedel, S.W. 1992, in Astronomical Data Analysis Software and Systems I, eds. D.M. Worrall, C. Biemesderfer, & J. Barnes (San Francisco:ASP), p. 379

Wright, E.L., *et al.* 1994, ApJ, 420, 1

Wright, E.L., *et al.* 1992, ApJ, 396, L13